\documentclass[aps prapplied,twocolumn,superscriptaddress, longbibliography, floatfix]{revtex4-1}

\usepackage{graphicx}
\usepackage{amsmath}
\usepackage{bm}
\usepackage{color,soul}
\usepackage{gensymb}

\definecolor{mygreen}{rgb}{0.0, 0.7, 0.2}

\begin{document}
\setlength{\textfloatsep}{12pt}

\title{Wavelength-accurate nonlinear conversion through wavenumber selectivity in photonic crystal resonators}


\author{Jordan R. Stone}
\email{jstone12@umd.edu}
\affiliation{Joint Quantum Institute, NIST/University of Maryland, College Park, MD 20742}
\affiliation{National Institute for Standards and Technology, Gaithersburg, MD 20899}

\author{Xiyuan Lu}
\affiliation{Joint Quantum Institute, NIST/University of Maryland, College Park, MD 20742}
\affiliation{National Institute for Standards and Technology, Gaithersburg, MD 20899}

\author{Gregory Moille}
\affiliation{Joint Quantum Institute, NIST/University of Maryland, College Park, MD 20742}
\affiliation{National Institute for Standards and Technology, Gaithersburg, MD 20899}

\author{Daron Westly}
\affiliation{National Institute for Standards and Technology, Gaithersburg, MD 20899}

\author{Tahmid Rahman}
\affiliation{Joint Quantum Institute, NIST/University of Maryland, College Park, MD 20742}
\affiliation{National Institute for Standards and Technology, Gaithersburg, MD 20899}

\author{Kartik Srinivasan}
\affiliation{Joint Quantum Institute, NIST/University of Maryland, College Park, MD 20742}
\affiliation{National Institute for Standards and Technology, Gaithersburg, MD 20899}


\date{\today}

\begin{abstract}
Integrated nonlinear wavelength converters transfer optical energy from lasers or quantum emitters to other useful colors, but chromatic dispersion limits the range of achievable wavelength shifts. Moreover, because of geometric dispersion, fabrication tolerances reduce the accuracy with which devices produce specific target wavelengths. Here, we report nonlinear wavelength converters whose operation is not contingent on dispersion engineering; yet, the output wavelengths are controlled with high accuracy. In our scheme, coherent coupling between counter-propagating waves in a photonic crystal microresonator induces a photonic bandgap that isolates (in dispersion space) specific wavenumbers for nonlinear gain. We first demonstrate the wide applicability of this strategy to parametric nonlinear processes, by simulating its use in third harmonic generation, dispersive wave formation in Kerr microcombs, and four-wave mixing Bragg scattering. In experiments, we demonstrate Kerr optical parametric oscillators in which such wavenumber-selective coherent coupling designates the signal mode. As a result, differences between the targeted and realized signal wavelengths are $<0.3$ percent. Moreover, leveraging the bandgap-protected wavenumber selectivity, we continuously tune the output frequencies by nearly $300$ GHz without compromising efficiency.  Our results will bring about a paradigm shift in how microresonators are designed for nonlinear optics, and they make headway on the larger problem of building wavelength-accurate light sources using integrated photonics. 
\end{abstract}


\maketitle


\section{\label{sec:intro}Introduction}
Controlling integrated microsystems to generate light with properties specifically geared to applications is a fundamental ambition of photonics research. For example, optical atomic clocks require ultra-coherent laser light with wavelengths precisely matched to atomic transitions, and future hybrid quantum networks will interface sources of nonclassical light (e.g., single photons) tuned to qubit wavelengths \cite{ludlow2015optical, bothwell2022resolving, wehner2018quantum, elshaari2020hybrid}. A powerful tool to meet the demands of such systems is optical nonlinearity, which can mold light on a quantum level and stimulate wavelength conversion (e.g., by four-wave mixing (FWM)) for spectral access beyond conventional laser gain. In particular, optical microresonators with Kerr ($\chi^{(3)}$) nonlinearity have, after multiple groundbreaking demonstrations, become a linchpin of nonlinear photonics. They support microcombs for frequency synthesis, timekeeping, and sensing \cite{spencer2018optical, newman2019architecture, riemensberger2020massively, gaeta2019photonic}; optical parametric oscillators ($\mu$OPOs) for wavelength-flexible sources of laser light \cite{lu2020chip, sayson2019octave,marty_photonic_2021}, squeezed light \cite{vernon2019scalable, dutt2015chip} and (when operated below threshold) entangled photon pairs \cite{kues2019quantum, lu2019chip}; four-wave mixing Bragg scattering (FWM-BS) for spectral translation of single photons \cite{li2016efficient}; third harmonic generation (THG) \cite{carmon_visible_2007, surya2018efficient}; and more. Although appreciable efficiencies have been shown in some cases, it remains a challenge to ensure \textit{a priori} (i.e., before testing) that a specific device will achieve the desired combination of wavelength accuracy and efficiency. 
\begin{figure*}[ht]
    \centering
    \includegraphics[width=425 pt]{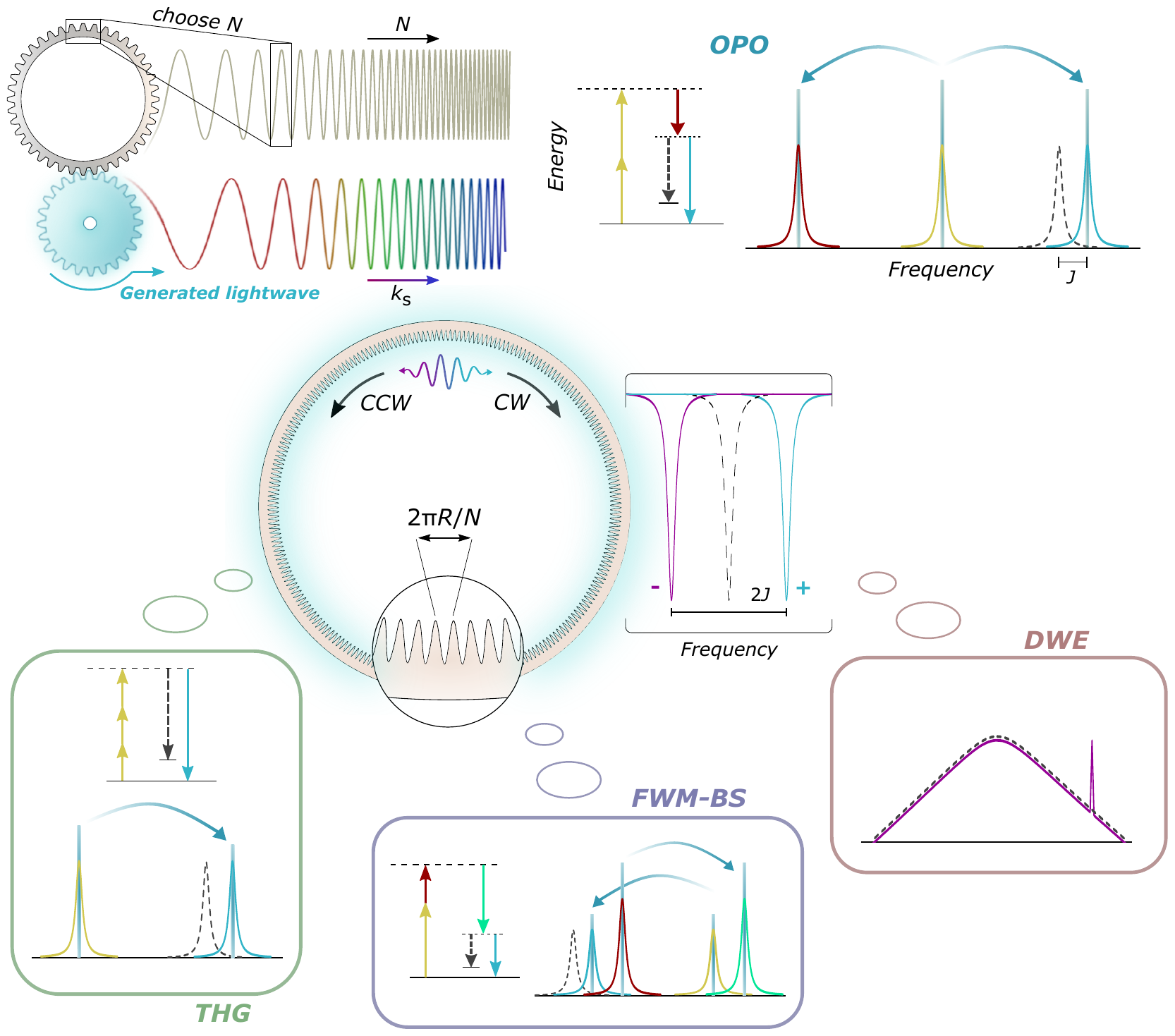}
    \caption{\textbf{Conceptual depictions of wavenumber-selective nonlinear wavelength conversion in Kerr photonic crystal microresonators.} Spatial modulation of the microresonator inner sidewall (pictured center) with a grating period $2\pi R/N$, where $N$ is an integer, coherently couples clockwise (CW) and counter-clockwise (CCW) travelling-wave modes with the azimuthal mode number $m_{\rm{s}}=N/2$. Coherent coupling induces a frequency splitting between two supermodes, denoted $`+'$ and $`-'$, with frequency separation $2J$, where $J$ is proportional to the sidewall modulation amplitude. We link the spatial frequency of sidewall modulation, $N$, to the wavenumber, $k_{\rm{s}}=N/2R$, of an output wave that is generated via nonlinear wavelength conversion. Hence, the photonic crystal resonator functions as a sort of gear, as illustrated in the upper left, to accurately control the wavelengths produced by a given device. Bottom portion: In resonators with normal group velocity dispersion (GVD), four-wave mixing (FWM) cannot occur between travelling-wave modes due to energy non-conservation (see energy level diagrams), but frequency matching can be realized using one of the supermodes. This allows, for example, optical parametric oscillation (OPO), third harmonic generation (THG), and FWM Bragg scattering (FWM-BS) in microresonators with purely normal GVD, and dispersive-wave enhancements (DWEs) in microresonators with purely anomalous GVD that support soliton microcombs.}
    \label{fig:one}
\end{figure*}

To elucidate the problem, we recall some basic design considerations for Kerr-nonlinear microresonators, focusing on commonly used microring devices. Fundamentally, energy and momentum conservation regulate FWM \cite{boyd2020nonlinear}; therefore, to within (approximately) a resonator linewidth, a set of resonator modes should obey:
\begin{subequations}\label{eq:cons}
\begin{align}
\sum_{i}\nu_{i} = \sum_{j}\nu_{j},\\
\sum_{i}m_{i} = \sum_{j}m_{j},    
\end{align}
\end{subequations}
where $m_{\rm{i}}$ is the azimuthal number (fundamentally related to the wavenumber) associated with a resonator mode with frequency $\nu_{\rm{i}}$, and left-hand (right-hand) terms denote photons created (annihilated) in the FWM process. Equation \ref{eq:cons} is exact when $\nu_{\rm{i}}$ and $m_{\rm{i}}$ refer to field quantities. In general, group velocity dispersion (GVD) induces a frequency mismatch, such that a set of modes satisfying Eq.~\ref{eq:cons}b does not simultaneously satisfy Eq.~\ref{eq:cons}a. The strategic `dispersion engineering' of modes to satisfy both parts of Eq.~\ref{eq:cons} is ubiquitous in guided-wave nonlinear photonics, with the most popular approach being to complement material dispersion with dispersion arising from the microresonator geometry \cite{yang2016broadband, okawachi2014bandwidth, lu2019efficient}. However, modeling broadband spectra, such as octave-spanning microcombs or $\mu$OPOs with widely-separated wavelengths, often requires retaining six or more orders in a Taylor expansion of $\nu_{\rm{i}}(m_{\rm{i}})$ around the pump wavelength \cite{yu2019tuning, briles2018interlocking}. In this regime, the mode wavelengths that satisfy Eq.~\ref{eq:cons} are extremely sensitive to geometry. Hence, small errors in the device geometry (arising from either fabrication uncertainties or incomplete modeling) can amount to significant differences between the simulated and experimentally-observed spectrum. This necessitates the fabrication of many (often, hundreds or more) devices with nanometer-scale parameter variations. Ultimately, one negotiates a trade-off between the number of devices that require testing and the dispersion tolerance of a given application. In many cases, a simple geometry-based solution to realize a particular GVD (e.g., one based on controlling the dimensions of a waveguide) does not exist. To make matters worse, unwanted nonlinear couplings (e.g., Raman scattering, mode competition, etc.) can compete with or even suppress the targeted process \cite{li2016efficient, stone2022conversion, zhang2021squeezed, tang2020widely}.

Here, we demonstrate Kerr-nonlinear wavelength conversion for which the $m$ values of participating resonator modes are guaranteed from design; yet, our method actually alleviates design constraints, naturally suppresses unwanted nonlinear couplings, and does not rely on sensitive control of higher-order GVD. We show how coherent coupling between counter-propagating waves in a photonic crystal microresonator induces controlled frequency splittings that balance the underlying GVD to satisfy Eq.~\ref{eq:cons}. We analyze $\mu$OPO, THG, dispersive-wave enhancement (DWE) in microcombs, and FWM-BS by introducing coherent coupling into simulations of those systems, and we prove our ideas experimentally using the flexible example of $\mu$OPO. Through the photonic crystal grating period, we dictate $m$ values for the signal modes in three different $\mu$OPOs, and we showcase their tolerance to higher-order GVD by reproducing the same signal wavelength when pumping four separate modes of a single device. Generated signal wavelengths agree with simulations to within $0.3$ \%. We characterize the $\mu$OPOs by their threshold power and conversion efficiency, and we find that our measurements agree with a model based on the Lugiato-Lefever Equation. Finally, we highlight the protected nature of our method by tuning the $\mu$OPO output frequencies continuously over $300$ GHz without sacrificing efficiency or inducing mode hopping. Our work re-envisions the design process for nonlinear wavelength converters, enables nonlinear optics in new spectral regions and with strongly-dispersive materials, and invites fundamental studies of nonlinear physics in photonic crystal microresonators. 

\section{Photonic crystal-mediated FWM featuring wavenumber selectivity}
Figure \ref{fig:one} depicts a photonic crystal microresonator and illustrates the four FWM processes we study. For concreteness, we consider silicon nitride (SiN) microrings where the ring width, $RW'$, varies along the inner boundary according to $RW'=RW+A_{\rm{mod}}\text{cos}(N\theta)$, where $RW$ is the nominal ring width, $N$ is an integer, and $\theta$ is the resonator azimuthal angle. Therefore, the spatial period of modulation is $2\pi R/N$, where $R$ is the ring radius. The modulation creates a refractive index grating that coherently couples clockwise (CW) and counter-clockwise (CCW) travelling-wave (TW) modes with the azimuthal number $m=N/2$, where $m$ is an integer related to the wavenumber, $k$, by $k=m/R$. Hence, we say the coherent coupling is ``wavenumber-selective." The coupling rate, $J$, is proportional to $A_{\rm{mod}}$ and corresponds to half the frequency splitting between two supermodes, denoted `+' and `-' for the higher- and lower-frequency resonances, respectively (pictured center). This type of resonator has numerous functionalities, including sensing \cite{urbonas2016air, lo2017photonic} and the slowing of light \cite{lu2022high}. In the context of nonlinear optics, pump mode hybridization has been used to induce spontaneous pulse formation and facilitate parametric oscillations in resonators with normal GVD. \cite{yu2021spontaneous, lu2022kerr, black2022optical}. Moreover, modulations with different $N$ values can be combined to realize multi-wavelength dispersion engineering \cite{lu2020universal, moille2022arbitrary, lucas2022tailoring}. In these experiments and others, $J$ could be made larger than the resonator free spectral range (FSR) without reducing the quality factor ($Q$).

In our experiments, we focus on $\mu$OPOs, which generate monochromatic signal and idler waves from a continuous-wave (CW) pump laser through resonantly-enhanced degenerate FWM, as shown at the top (energy diagram and optical spectrum) of Fig.~\ref{fig:one}. Momentum conservation requires $2m_{\rm{p}}=m_{\rm{s}}+m_{\rm{i}}$, where $m_{\rm{p}}$, $m_{\rm{s}}$, and $m_{\rm{i}}$ are azimuthal numbers for the pump, signal, and idler modes, respectively. Hence, mode pairs with $m=m_{\rm{p}}\pm \mu$, where $\mu$ is an integer, may support $\mu$OPO if their resonance frequencies obey Eq.~\ref{eq:cons}a. In general, GVD prevents such frequency matching; \textit{i.e.}, the associated FWM process does not conserve energy. In Fig.~\ref{fig:one}, gray dashed lines in the energy diagrams and optical spectra illustrate how GVD suppresses FWM. To quantify this concept, we define the frequency mismatch as:
\begin{equation}\label{eq:mis}
\Delta \nu = \nu_{\rm{\mu}}+\nu_{\rm{-\mu}}-2\nu_{\rm{0}}, 
\end{equation}
where $\nu_{\rm{0}}$ is the pump mode frequency, and $\nu_{\rm{\mu}}$ is the mode frequency associated with the azimuthal number $m_{\rm{p}}+\mu$. Normal GVD gives $\Delta \nu < 0$ for all $\mu$ and thus prevents FWM. Nonetheless, applying an appropriate shift to $\nu_{\rm{\mu}}$ (or $\nu_{\rm{-\mu}}$) will restore energy conservation and activate the $\mu$OPO, as illustrated by the blue lines in Fig. ~\ref{fig:one}. We can realize this shift via the `+' supermode; changing to the `+' basis gives the transformation:
\begin{equation}\label{eq:good}
        \Delta \nu_{\rm{+}} =
        \begin{cases}
        \Delta \nu_{\rm{CW}}+J,& m=N/2\\
        \Delta \nu_{\rm{CW}},& \text{else}
        \end{cases}
\end{equation}
where $\Delta \nu_{\rm{CW}}$ is the frequency mismatch in the CW basis. Hence, we select $m_{\rm{s}}$ by choosing $N=2m_{\rm{s}}$, and the $\mu$OPO is activated when $J=-\Delta \nu_{\rm{CW}}$.

Importantly, coherent coupling in photonic crystal resonators can facilitate other FWM processes besides $\mu$OPO, as illustrated in Fig.~\ref{fig:one}. Specifically, we explore THG, FWM-BS, and DWE, all of which involve wide spectral gaps between their constituent wavelengths and thus exhibit $\Delta \nu$ spectra that are difficult to control exclusively via the microresonator cross-sectional geometry. In each case, we can re-define $\Delta \nu$ according to Eq.~\ref{eq:cons}a (see Appendix A) and employ coherent coupling to restore energy conservation by balancing $\Delta \nu_{\rm{CW}}$ with $J$. In Fig.~\ref{fig:one}, energy diagrams and optical spectra show how shifting the frequency of one mode can promote THG and FWM-BS. The DWE process merits special elaboration. Bright soliton microcombs operate in a regime of anomalous GVD, but certain wavelengths with normal GVD can exhibit local power enhancements (i.e., DWE) \cite{briles2018interlocking, moille2021tailoring}. The DWE phenomenon is useful to aid self-referencing, but the dispersive-wave (DW) wavelengths are difficult to control due to their reliance on higher-order GVD. We envision using wavenumber-selective coherent coupling to dictate the $m$ values of DWs. Because of the underlying anomalous GVD, DWs would be resonant with the `-' supermode. This scheme could operate without tailoring higher-order GVD and deterministically select harmonic wavelengths for self-referencing, thus augmenting microcombs spectrally-tailored with Fourier synthesis \cite{moille2022arbitrary}.  

\begin{figure}[ht]
    \centering
    \includegraphics[width=245 pt]{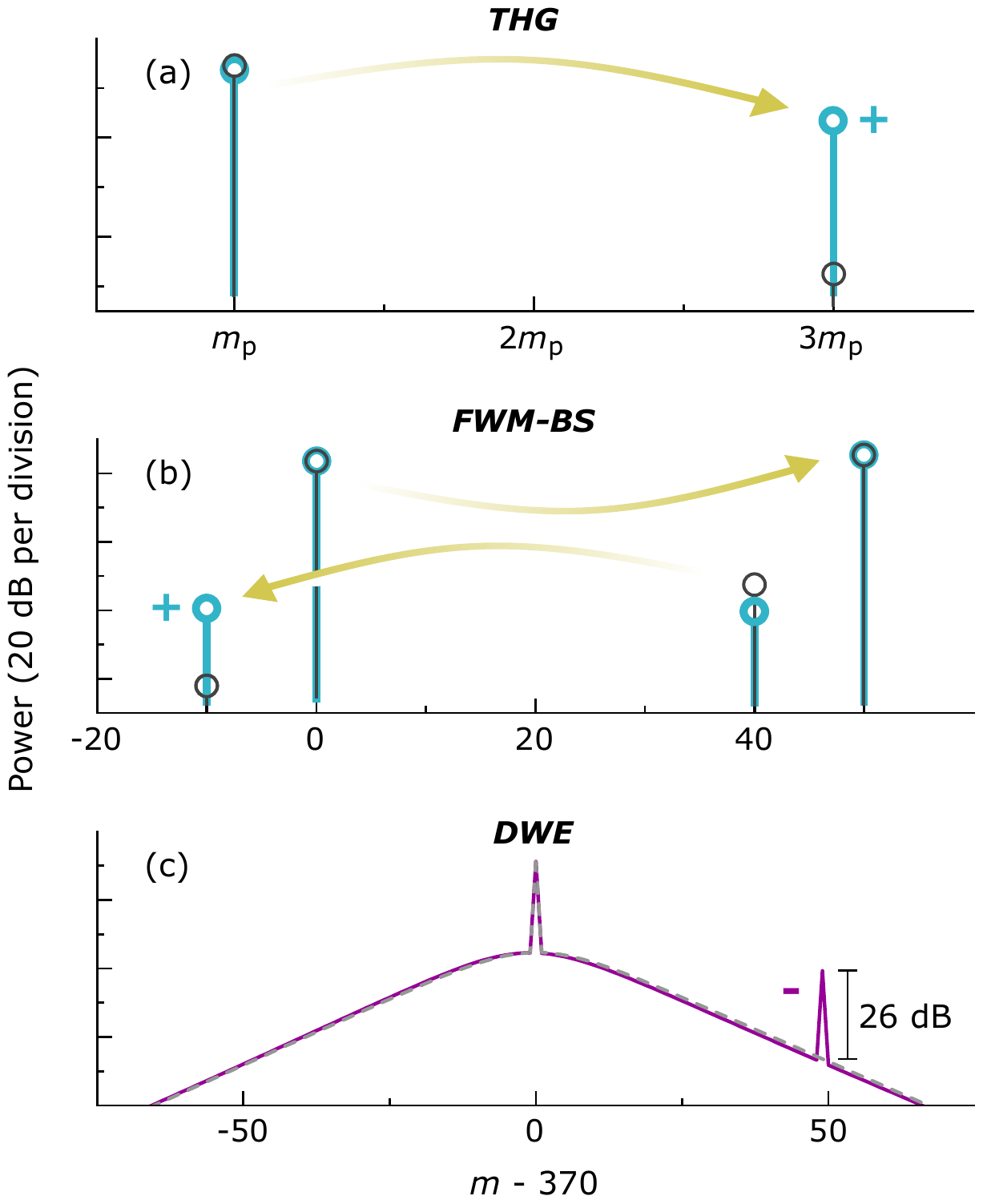}
    \caption{\textbf{Simulations of nonlinear wavelength conversion in Kerr photonic crystal microresonators.} The $m$ values designated for coherent coupling are marked by a blue `+' or a purple `-', depending on which supermode is utilized. (a) Simulated THG spectrum, both with (blue) and without (gray) photonic crystal-mediated coherent coupling ($J$). The simulation parameters are $\Delta \nu_{\rm{CW}} = 12.5$ GHz, $J=12.425$ GHz (blue data only), and $P_{\rm{in}}=250$ $\mu$W. (b) Simulated FWM-BS spectrum, both with (blue) and without (gray) coherent coupling. The simulation parameters are $D_{\rm{2}}/2\pi = -25$ MHz per mode, which corresponds to $\Delta \nu_{\rm{CW}} = 12.5$ GHz, $J=12.6$ GHz (blue data only), and $P_{\rm{in}}=5$ mW for both pump lasers. (c) Simulated Kerr microcomb spectrum with (purple) and without (dashed gray) coherent coupling. Coherent coupling is used for dispersive wave enhancement (DWE), to increase the power of a single microcomb tone by 26 dB. The simulation parameters are $D_{\rm{2}}/2\pi=10$ MHz per mode,  $J=13.75$ GHz (purple data only), and $P_{\rm{in}}=15$ mW. Definitions of $\Delta\nu$ for THG and FWM-BS are given in Appendix A.}
    \label{fig:two}
\end{figure}

To prove our ideas, we analyze THG, FWM-BS, and DWE in resonators with either purely normal (for THG and FWM-BS) or purely anomalous (for DWE) GVD by including coherent coupling in simulations of those systems. We reserve $\mu$OPO simulations for the next section, where we aim to verify our model with experiments. We use a set of coupled-mode equations (CMEs) to simulate THG, and a pair of coupled Lugiato-Lefever Equations (LLEs) to simulate FWM-BS and DWE (for details, see Appendix A). Importantly, we include the coherent coupling explicitly in our models; i.e., we do not manually insert frequency shifts into the GVD, since this would not account for the hybridization of CW/CCW modes. We define the mode spectra and perform simulations in the CW/CCW basis. To include coherent coupling, we allow one CW mode to exchange energy with its CCW counterpart at a coupling rate $J$ that is continuously tunable. In Fig. \ref{fig:two}, we present simulated optical spectra for THG, FWM-BS, and DWE. The gray data correspond to simulations with $J=0$, while blue or purple data (when utilizing the `+' or `-' supermodes, respectively) correspond to simulations where $J$ is tuned to maximize the signal (or DW) power. 

In our simulations, we assign to all modes a (critically-coupled) loaded linewidth $\kappa/2\pi=500$ MHz. In THG simulations, we set $\Delta\nu_{\rm{CW}}=12.5$ GHz and $P_{\rm{in}}=250$ $\mu$W, where $P_{\rm{in}}$ is the pump power. This $P_{\rm{in}}$ value efficiently drives THG but is below the saturation power (see Appendix A). We apply coherent coupling to the third-harmonic mode. When $J=0$, the third harmonic power, $P_{3H}\approx 2.7$ nW. We find that $J=12.425$ GHz maximizes $P_{3H}$, in accordance with Eq. \ref{eq:good}, increasing it to $P_{3H}\approx 3$ $\mu$W, as shown in Fig. \ref{fig:two}a. 

To model FWM-BS, we simulate a microresonator pumped by two separate pump lasers resonant with modes $m=370$ and $m=420$. $P_{\rm{in}}=5$ mW for both pump lasers. A low-power input seed, resonant with mode $m=410$, is also injected into the resonator. FWM-BS converts input seed photons to output signal photons resonant with $m=360$. We set $D_{2}/2\pi=-25$ MHz per mode, where $D_{2}$ is the second-order term in a Taylor series expression of the integrated dispersion, $D_{\rm{int}}=\nu_{\mu}+(\nu_{\rm{0}}-\mu \text{FSR})$. This $D_{2}$ value corresponds to $\Delta \nu_{\rm{CW}}=12.5$ GHz. We apply coherent coupling to the signal mode. When $J=0$, virtually no seed photons are converted. When $J=12.6$ GHz, $\approx 25~\%$ of input photons are converted to wavelength-shifted output photons, as shown in Fig. \ref{fig:two}b. Notably, Liu \textit{et al.} recently proposed a dispersion engineering approach to FWM-BS that is also based on coherent coupling between CW/CCW modes \cite{liu2021tunable}. 

\begin{figure}[ht]
    \centering
    \includegraphics[width=245 pt]{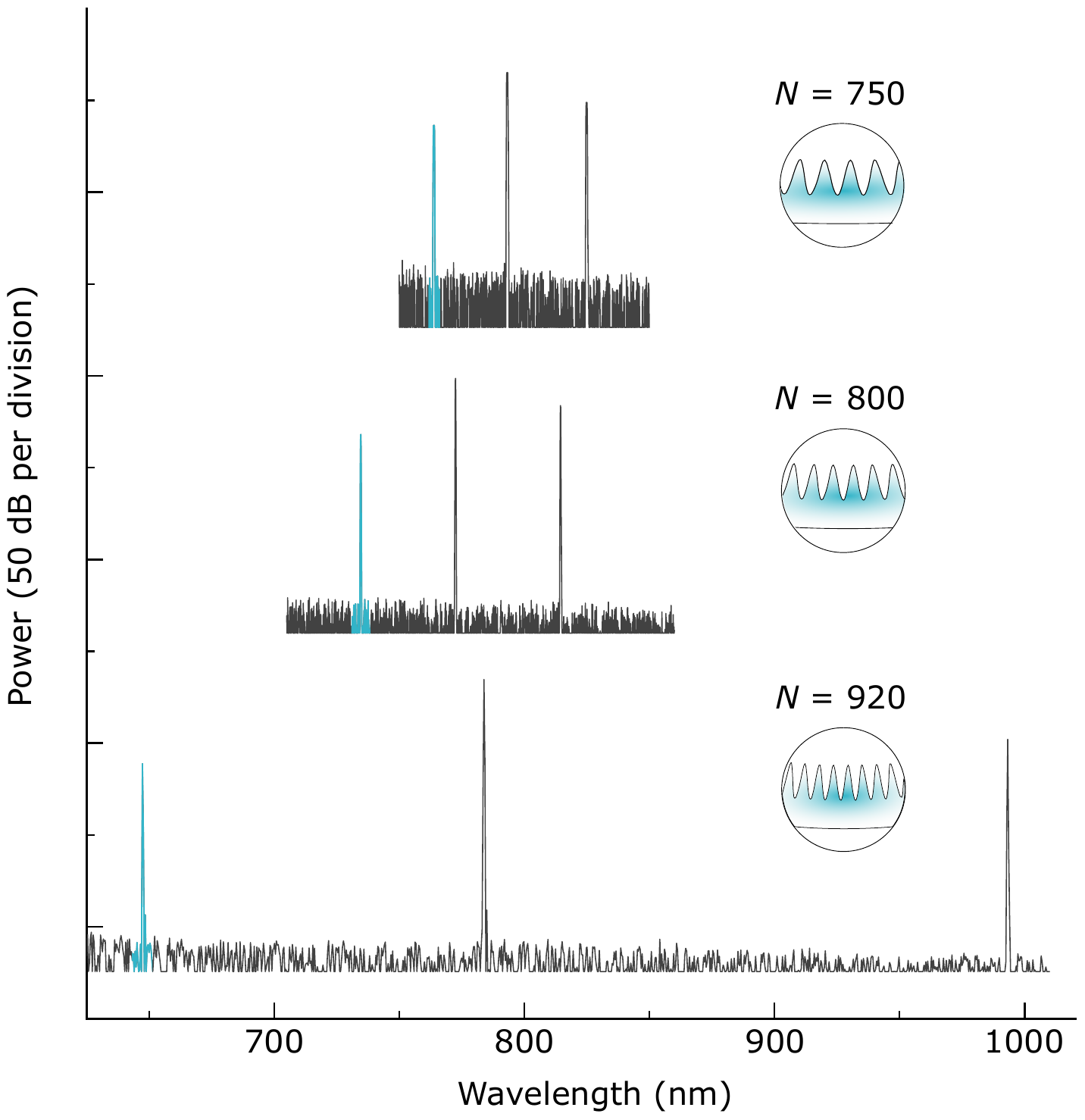}
    \caption{\textbf{Wavenumber-selective $\mu$OPO in Kerr photonic crystal microresonators.} Optical spectra generated in three $\mu$OPO devices. From top to bottom, $N=(750, 800, 920)$, and $A_{\rm{mod}}=(5, 10, 25)$ nm is chosen to balance the underlying GVD. In each spectrum, the line corresponding to the signal wave is colored blue, and the signal mode number, $m_{\rm{s}}=N/2$. Every device exhibits normal GVD at the pump, signal, and idler wavelengths.}
    \label{fig:three}
\end{figure}

\begin{figure*}[ht]
    \centering
    \includegraphics[width=450 pt]{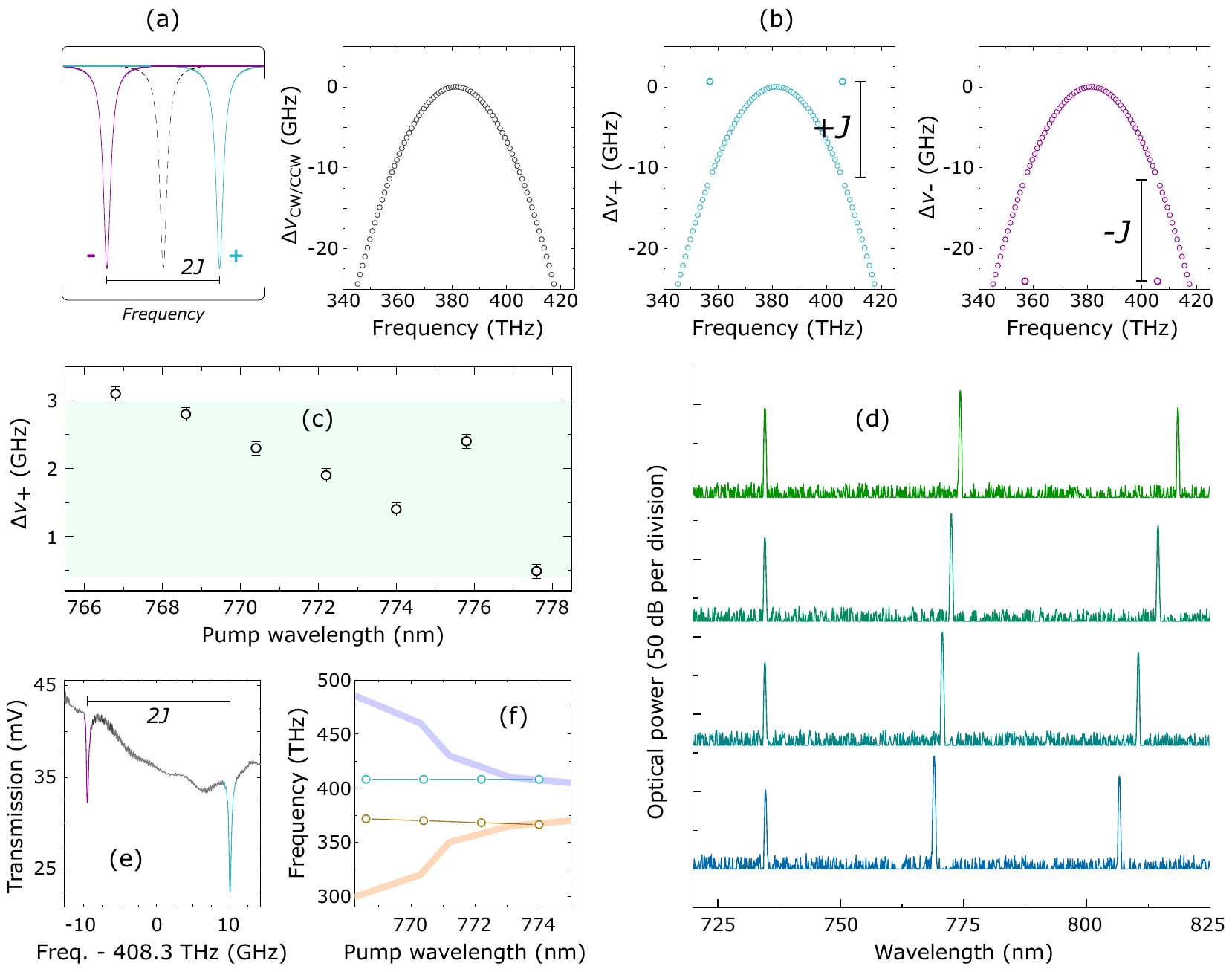}
    \caption{\textbf{Optical parametric oscillation using selective splitting in undulated microresonators (OPOSSUM).} (a) Conceptual transmission spectrum illustrating the frequency splitting of a travelling-wave mode (gray dashed line) into two standing-wave supermodes with frequency separation $2J$. (b) Simulated $\Delta\nu$ spectra of an OPOSSUM device in the CW/CCW basis (left), the `+' basis (center), and the `-' basis (right). In the `+' basis, a single mode pair is frequency matched to allow FWM, and normal GVD mismatches all other mode pairs. (c) $\Delta\nu_{\rm{+}}$ versus pump wavelength for an OPOSSUM device with $R=25$ $\mu$m, $RW=925$ nm, $H=600$ nm, and $N=800$. Vertical error bars correspond to the range in $\Delta\nu_{\rm{+}}$ values obtained when the measurement is repeated many ($\approx 10$) times. The pale green stripe indicates $\Delta\nu_{\rm{+}}$ values conducive to $\mu$OPO. (d) Optical spectra obtained from pumping four different modes (with wavelengths between $768$~nm to $774$~nm) in the OPOSSUM device. (e) Transmission spectrum of the same device showing '+' and '-' supermodes (blue and purple, respectively) with frequency separation $2J \approx 20$ GHz. (f) OPOSSUM signal (blue circles) and idler (gold circles) frequencies versus pump wavelength. The pale stripes show the same data, taken from Ref. \cite{lu2020chip}, for a device without coherent coupling that relies on higher-order GVD engineering for frequency matching.}
    \label{fig:four}
\end{figure*}

To simulate DWE, we set $D_{2}/2\pi=10$ MHz per mode and apply coherent coupling to the $m=419$ mode. A laser, resonant with mode $m=370$, pumps the resonator with $P_{\rm{in}}=15$ mW. When $J=0$, the microcomb spectrum exhibits a smooth sech$^2$ profile with no DWEs. When $J=13.75$ GHz, we observe a $26$ dB power enhancement at the targeted mode, as shown in Fig. \ref{fig:two}c. In Appendix A, we characterize our simulations in more detail. Remarkably, our modeling captures wavelength conversion into the supermodes, thus illustrating the applicability of our scheme to a variety of Kerr-nonlinear processes. 

To validate the main elements of our approach in experiments, we choose an additional Kerr-nonlinear process, that of degenerately-pumped $\mu$OPO. In processes like THG and FWM-BS, the potential output wavelength is known \textit{a priori} from the input wavelengths, with the efficiency of conversion depending on $\Delta\nu$ (as well as other parameters not dependent on the phase- and frequency-matching strategy, namely, resonator-waveguide coupling \cite{li2016efficient}).  In contrast, the $\mu$OPO output wavelengths are not determined solely by the input wavelengths, but can widely vary depending on GVD. Therefore, $\mu$OPOs provide an ideal experimental test of wavenumber-selective FWM.

To this end, we perform experiments that demonstrate \textit{a priori} control over $m_{\rm{s}}$ in $\mu$OPO devices with $N=2m_{\rm{s}}$. In Fig. \ref{fig:three}, we present optical spectra generated in three different photonic crystal microresonators with $RW'$ modulations parameterized by $N=(750, 800, 920)$ and $A_{\rm{mod}}=(5, 10, 25)$ nm. In each device, $A_{\rm{mod}}$ is chosen to balance the underlying normal GVD (in section \ref{sec:OPOSSUM}, we explain our design process in more detail). We pump a fundamental transverse-electric (TE0) resonator mode near $780$ nm, and we observe one of two outcomes: a $\mu$OPO with $m_{\rm{s}}=N/2$ when $J$ compensates for $\Delta \nu_{\rm{CW}}$ (i.e., the three spectra in Fig.~\ref{fig:three}), or a CW state (i.e., no wavelength conversion; data not shown in Fig.~\ref{fig:three}) preserved by normal GVD and an incommensurate balance of $\Delta \nu_{\rm{CW}}$ and $J$. We confirm the $m_{\rm{s}}$ values from mode transmission spectroscopy, and we measure (simulate) signal wavelengths of $763.5$ nm ($761.5$ nm), $735$ nm ($735.8$ nm), and $648$ nm ($649.9$ nm). This binary distribution of measurement outcomes affirms the protected nature of wavelength conversion in our experiments.

\section{OPOSSUM}\label{sec:OPOSSUM}
\begin{figure*}[ht]
    \centering
    \includegraphics[width=400 pt]{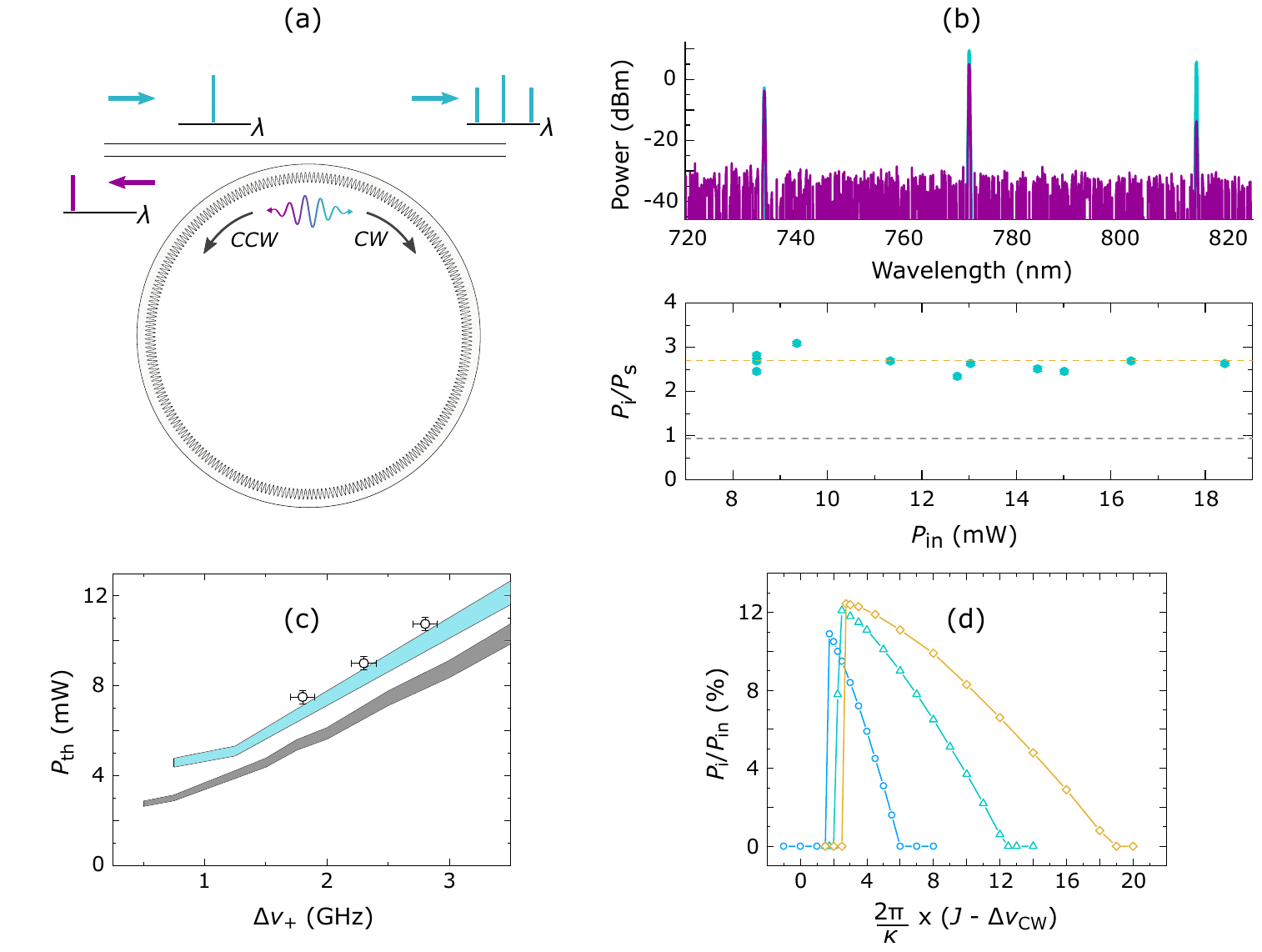}
    \caption{\textbf{Modeling OPOSSUM with the Lugiato-Lefever Equation (LLE).} (a) Illustration of input and output spectra from an OPOSSUM device. Due to coherent coupling between CW and CCW waves in the signal mode, a fraction of signal photons are outcoupled in a direction that is counter-propagating to the injected pump. (b) Top panel: A sample OPOSSUM spectrum calibrated to indicate the on-chip power. Blue data correspond to transmitted light (i.e., light that is outcoupled in a direction co-propagating with the pump laser), and purple data correspond to reflected light. Bottom panel: Measured ratio ($P_{\rm{i}}/P_{\rm{s}}$) of transmitted idler power ($P_{\rm{i}}$) to transmitted signal power ($P_{\rm{s}}$) versus $P_{\rm{in}}$. The orange (gray) dashed line is a theoretical prediction based on LLE simulations that include (do not include) coherent coupling. (c) Measured threshold power, $P_{\rm{th}}$, versus $\Delta\nu_{\rm{+}}$. Vertical error bars are due to uncertainties in optical losses between the input and output fibers, calculated as one standard deviation in loss measurements performed for many separate devices. Horizontal error bars correspond to the range in $\Delta\nu_{\rm{+}}$ values obtained when the measurement is repeated many ($\approx 10$) times. The blue and gray stripes are theoretical predictions based on LLE simulations, with (blue) and without (gray) coherent coupling; i.e., the gray stripe is derived from an LLE where $\Delta\nu_{\rm{CW}}$ is adjusted to realize frequency matching. The finite thicknesses of theory curves correspond to uncertainties in the value of the Kerr nonlinear coefficient. (d) Simulated idler conversion efficiency, $P_{\rm{i}}/P_{\rm{in}}$, versus normalized $J$ for $P_{\rm{in}} = 10$ mW (blue circles), $P_{\rm{in}}=20$ mW (green triangles), and $P_{\rm{in}}=30$ mW (gold diamonds).}
    \label{fig:five}
\end{figure*}
We now explain our procedures for designing photonic crystal microresonators and testing them post-fabrication (for details about the fabrication process, see Appendix B). We refer to the $\mu$OPO mechanism as OPOSSUM, which stands for optical parametric oscillation using selective splitting in undulated microresonators. To start, we reiterate the impact of wavenumber-selective coherent coupling on the resonator mode spectrum: CW and CCW modes with $m=N/2$ hybridize into two supermodes with frequency separation $2J$, as illustrated in Fig. \ref{fig:four}a. Hence, OPOSSUM devices exhibit three $\Delta \nu$ spectra, denoted $\Delta \nu_{\rm{CW/CCW}}$, $\Delta \nu_{\rm{+}}$, and $\Delta \nu_{\rm{-}}$, depending on the basis used. To choose values for $RW, N$, and $A_{\rm{mod}}$ (the SiN thickness, $H$, is fixed by our current stock of SiN, and $R=25~\mu$m), we simulate mode spectra using the finite-element method for devices without $RW'$ modulation. We calculate $\Delta \nu_{\rm{CW}}$ according to Eq. \ref{eq:mis} and choose a $RW$ value that exhibits broadband normal GVD. Then, we identify a target signal wavelength (e.g., $760$ nm, $735$ nm, and $650$ nm for the three devices related to Fig.~\ref{fig:two}b) and choose $N$ accordingly. To select $A_{\rm{mod}}$, we fabricate a set of devices with variations in $RW$, $A_{\rm{mod}}$, and $N$, and we measure the frequency splittings to calibrate $J(N, RW, A_{\rm{mod}})$. Using our calibrations, we set $A_{\rm{mod}}$ for a particular device to balance $\Delta\nu_{\rm{CW}}$. Figure \ref{fig:four}b depicts simulated $\Delta \nu_{\rm{CW/CCW}}$, $\Delta \nu_{\rm{+}}$, and $\Delta \nu_{\rm{-}}$ spectra for a device with $RW=925$ nm, $H=600$ nm, and $N=800$. Notably, the $\Delta \nu_{\rm{+}}$ spectrum is discontinuous at the signal and idler frequencies, where $\Delta \nu_{\rm{+}}=\Delta \nu_{\rm{CW}}+J$. This suggests that OPOSSUM suppresses FWM involving modes other than the targeted signal and idler, since at these frequencies the resonator exhibits strong normal dispersion.

Next, we perform experiments to characterize OPOSSUM. We fabricate the OPOSSUM device simulated in Fig.~\ref{fig:four}b and measure the TE0 mode wavelengths to calculate $\Delta \nu_{\rm{+}}$. Importantly, $\Delta \nu_{\rm{+}}$ depends on $m_{\rm{p}}$; hence, tuning the pump wavelength can correct for fabrication uncertainties and, more generally, ensure reliable operation. To concretize this idea, we measure $\Delta \nu_{\rm{+}}$ versus pump wavelength, as shown in Fig.~\ref{fig:four}c. We find that $\Delta \nu_{\rm{+}}$ decreases with increasing pump wavelength, with an exception near $776$ nm, where we observe mode crossings at the pump and idler wavelengths. In principle, we can generate a $\mu$OPO using any pump mode such that $\Delta \nu_{\rm{+}}>0$, provided $P_{\rm{in}}$ is large enough to induce compensating nonlinear mode-frequency shifts~\cite{stone2022conversion}. Realistically, however, we prefer $\Delta \nu_{\rm{+}}<3$ GHz. Greater $\Delta \nu_{\rm{+}}$ values require $P_{\rm{in}}>50$ mW to produce appreciable signal and idler powers; at this level, absorption-induced temperature shifts can destabilize the $\mu$OPO. At the same time, we require $\Delta \nu_{\rm{+}}>\kappa/4\pi$. In our example OPOSSUM device, the four pump modes spanning wavelengths $768$~nm to $774$~nm satisfy these requirements, as indicated by the pale stripe in Fig.~\ref{fig:four}c. Indeed, pumping any of these modes results in a $\mu$OPO. We record the optical spectra and present them in Fig.~\ref{fig:four}d. As expected, $m_{\rm{s}}$ is fixed - its value is protected by the wavenumber-selective coherent coupling, with an example transmission spectrum shown in Fig.~\ref{fig:four}e. In Fig.~\ref{fig:four}f, we present measurements of the signal and idler frequencies, $\nu_{\rm{s}}$ and $\nu_{\rm{i}}$, respectively, versus pump wavelength. We overlay similar data (pale stripes), taken from Ref.~\cite{lu2020chip}, for a $\mu$OPO system that relies on higher-order GVD, where the dispersion sensitivity is apparent from the large shifts in $\nu_{\rm{s}}$ (and $\nu_{\rm{i}}$) when tuning the pump laser between adjacent pump modes (i.e., with consecutive $m_{\rm{p}}$ values). By comparison, OPOSSUM is a robust mechanism for targeting specific wavelengths. 

\begin{figure}[ht]
    \centering
    \includegraphics[width=240 pt]{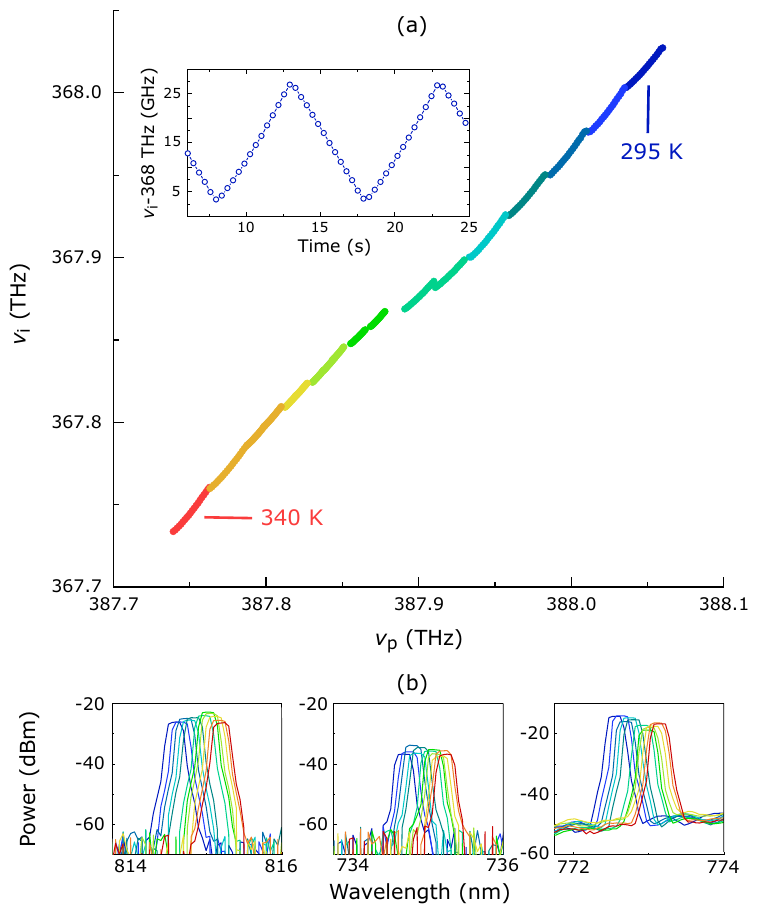}
    \caption{\textbf{Exploring wavelength tunability in OPOSSUM.} (a) Wavemeter measurement of $\nu_{\rm{i}}$ versus $\nu_{\rm{p}}$ at $11$ different temperatures (corresponding to the $11$ different colors). The temperature is used to coarsely tune $\nu_{\rm{i}}$, while controlling $\nu_{\rm{p}}$ enables fine tuning. Inset: Wavemeter measurement of $\nu_{\rm{i}}$ versus time during a $\nu_{\rm{p}}$ sweep. (b) Optical spectra zoomed into the idler, signal, and pump bands at each temperature (left, center, and right panels, respectively). These measurements show that output power is maintained across the tuning range.}
    \label{fig:six}
\end{figure}
Next, we investigate the OPOSSUM efficiency and threshold behavior. To model OPOSSUM, we simulate a pair of coupled LLEs that describe the intraresonator evolution of CW and CCW fields (see Appendix A for details). We are especially interested in connections between our experimental parameters and the power generated in signal and idler waves. Intuitively, we expect the signal wave, which occupies the `+' supermode, to propagate in both CW and CCW directions; hence, we should detect some signal light at the input (reflection) port of a device, as shown in Fig.~\ref{fig:five}a. In simulations, we observe approximately $20$ percent more signal power in the reflection port than the transmission port. This distribution is approximately independent of $P_{\rm{in}}$ and $\Delta \nu_{\rm{+}}$. In experiments, we measure an approximately equal distribution of signal power to the two ports. The top panel of Fig.~\ref{fig:five}b shows optical spectra calibrated to estimate the on-chip power levels at the transmission (blue) and reflection (purple) ports of the OPOSSUM device characterized in Fig.~\ref{fig:five}. The presence of reflected pump and idler light is due to Fresnel reflections at the waveguide facets, but such light is still strongly suppressed relative to the transmission port (e.g, $\approx$20~dB for the idler). Ultimately, large optical losses that occur during propagation from the reflection port to the optical spectrum analyzer prevent a precise measurement of the signal power distribution. A more precise comparison can be made between the transmitted powers of the signal and idler waves, denoted $P_{\rm{s}}$ and $P_{\rm{i}}$, respectively. Specifically, we calculate $P_{\rm{i}}/P_{\rm{s}}$ versus $P_{\rm{in}}$ and indicate our measurements with blue data points in the bottom panel of Fig.~\ref{fig:five}b. Our measurements agree with simulation results shown by the orange dashed line. Notably, we find that $P_{\rm{i}}/P_{\rm{s}}$ does not depend on $P_{\rm{in}}$. Moreover, the unequal distribution of photons between signal and idler waves is unique - previous (non-OPOSSUM) $\mu$OPO systems exhibited an equal distribution of photons ensured by the symmetry of degenerate FWM \cite{stone2022conversion}. In OPOSSUM, this symmetry is broken by CW/CCW coupling. Finally, we note that signal light propagating in the CW/CCW directions can be coherently re-combined outside the resonator to increase $P_{\rm{s}}$.   

To further characterize OPOSSUM, we measure the threshold power for parametric oscillation, $P_{\rm{th}}$, which is another important parameter of $\mu$OPO systems. Conveniently, we can measure $P_{\rm{th}}$ versus $\Delta \nu_{\rm{+}}$ by choosing different pump modes, as shown in Fig.~\ref{fig:five}c. The $P_{\rm{th}}$ values predicted from our model are shown by the blue stripe, and the $P_{\rm{th}}$ values predicted from a crude model (consisting of a single LLE wherein we shift the signal mode frequency by $J$) are shown by the gray stripe. Our measurements support the validity of our model. 
Next, we explore the robustness of OPOSSUM with respect to variations in $J$. Such an investigation conveys the design tolerance, i.e., the allowable errors in device geometry that can arise from fabrication uncertainties, of OPOSSUM. Specifically, we simulate OPOSSUM and calculate the conversion efficiency, $P_{\rm{i}}/P_{\rm{in}}$, versus $J$ for $P_{\rm{in}}=10, 20$, and $30$ mW, as shown in Fig.~\ref{fig:five}d. We find that the maximum conversion efficiency is $12.5$ percent for a critically-coupled resonator, which is the same result recently derived for other $\mu$OPO systems (the maximum conversion efficiency can be increased by overcoupling the resonator, at the cost of greater $P_{\rm{th}}$). Moreover, the range of $J$ values that supports a given efficiency increases with $P_{\rm{in}}$. For instance, to realize $P_{\rm{i}} \ge 2$ mW with $P_{\rm{in}}=20$ mW, we find $22 \le 2J \le 25$ GHz, where $\kappa/2\pi=500$ MHz and $\Delta \nu_{\rm{CW}} = 10$ GHz. For the device characterized in Figs.~\ref{fig:five}b-c, this corresponds roughly to $11$ nm $\le A_{\rm{mod}} \le 12.5$ nm. The possibility of increasing design tolerances using, e.g., temperature tuning, requires further study. 

Finally, we explore the wavelength tunability of OPOSSUM using the same device characterized in Figs.~\ref{fig:four} and \ref{fig:five}. Such tunability is of practical importance to nonlinear wavelength converters aiming for, e.g., specific atomic transitions. In our experiments, we sweep $\nu_{\rm{p}}$ by $\approx 25$ GHz in $5$ seconds while sustaining a $\mu$OPO, and we observe the resulting changes to $\nu_{\rm{i}}$ using a wavemeter ($\nu_{\rm{s}}$ can be inferred from $\nu_{\rm{i}}$ and $\nu_{\rm{p}}$ using Eq.~\ref{eq:cons}a). An example of these data is shown in the inset to Fig.~\ref{fig:six}a. We find $\frac{d\nu_{\rm{i}}}{d\nu_{\rm{p}}}\approx 1$. To extend the wavelength access of our OPOSSUM device, we increase its temperature, $T$, according to $\frac{d\nu_{\rm{0}}}{dT}\approx 4$ GHz/K and repeat the $\nu_{\rm{p}}$ sweep while recording $\nu_{\rm{i}}$. Figure \ref{fig:six}a shows our results from repeating this measurement at $11$ different temperatures (corresponding to the $11$ different colors in Fig.~\ref{fig:six}), from $T \approx 295$ K to $T \approx 340$ K, chosen to access all frequencies between $367.73 \le \nu_{\rm{i}} \le 368.02$ THz. (At some temperatures, we found that $\nu_{\rm{p}}$ could be swept $>25$ GHz while sustaining the $\mu$OPO. This is why some colors comprise more frequencies than others in Fig.~\ref{fig:six}a). At each temperature, we record the optical spectrum, as shown in Fig.~\ref{fig:six}b where we have magnified the idler, signal, and pump bands in the left, center, and right panels, respectively. Importantly, the $\mu$OPO output power is maintained across the entire tuning range. Moreover, the nearly $300$ GHz of tuning reported here was limited by instabilities in our setup at the higher temperatures. Given such stability, we expect that greater tuning ranges, possibly exceeding the FSR, are attainable. Our measurements suggest that a suitable choice of $N$, combined with continuous tunability, gives deterministic wavelength control with high accuracy. 

\section{Discussion}

One subtle and surprising aspect of our results is that `+' and `-' supermodes, which are standing-wave (and thus, momentum-less) superpositions of counter-propagating TW modes, can participate in FWM with TW modes. To conserve momentum, created signal photons should co-propagate with the pump laser, but these photons are off-resonant from the TW mode (strictly speaking, the TW modes do not exist at the signal wavelength. More precisely, we mean that such photons do not obey the boundary conditions for resonance). We hypothesize that, because $J\gg\kappa$, TW signal photons are created, but they are subsequently converted to the appropriate supermode more quickly than they are dissipated. In the future, we plan to analyze these dynamics in more depth. 

Importantly, through the OPOSSUM mechanism we achieve greater than $99.7$ \% wavelength accuracy without iterating fabrication runs (i.e., to target specific wavelengths, we identify $N$ values based only on our finite-element simulations, with little guidance from previous measurements). Moreover, temperature tuning beyond the $\approx 50$ K range we achieve in experiments will compensate for wavelength inaccuracies. In cases where $\Delta \nu_{\rm{+}}$ depends on $T$, one can leverage the relationship between $\Delta \nu_{\rm{+}}$ and $m_{\rm{p}}$. For instance, if $T$ must be adjusted so much that a $\mu$OPO is destabilized when pumping mode $m_{\rm{p}}$, then switching to $m_{\rm{p}}\pm1$ (depending on whether $T$ has been increased or decreased) will restore frequency matching. 

In conclusion, we have shown that coherent coupling in photonic crystal resonators can facilitate FWM-based nonlinear wavelength conversion without higher-order GVD. Moreover, proper design of the photonic crystal structure gives unprecedented control over signal wavelengths while protecting the FWM process from unwanted nonlinear couplings. To affirm simulation results that covered a wide range of $\chi^{(3)}$-nonlinear processes, we generated $\mu$OPOs with signal wavenumbers defined by the photonic crystal grating period. We measured the conversion efficiencies and threshold powers for multiple devices, and our measurements agreed with simulations. Finally, we demonstrated continuous tunability of the $\mu$OPO spectrum. Importantly, coherent coupling can be implemented in $\chi^{(2)}$-nonlinear systems, in addition to the $\chi^{(3)}$ systems discussed here. The devices and methods introduced here will be invaluable to future nanotechnologies that leverage application-tuned and wavelength-accurate nonlinear photonics.

\appendix
\section{Modeling nonlinear wavelength conversion in photonic crystal microresonators}
\begin{figure*}[ht]
    \centering
    \includegraphics[width=450 pt]{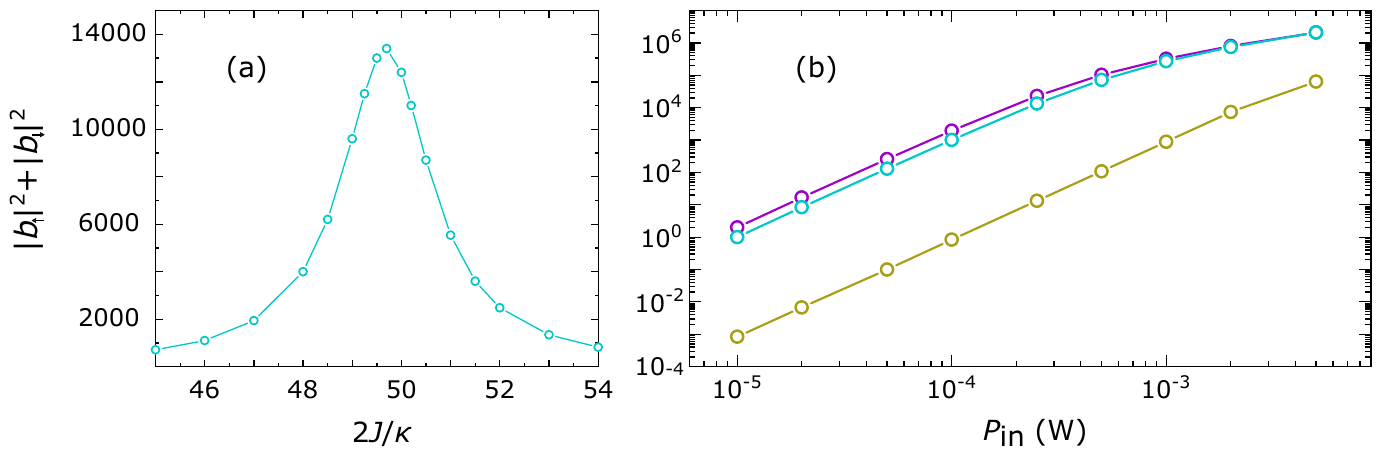}
    \caption{\textbf{Simulation analysis of THG in Kerr photonic crystal microresonators.} (a) Third harmonic power, $P_{3H}$, versus normalized coupling rate, $J$, for $P_{\rm{in}}=250$ $\mu$W and $\Delta\nu=50\times\kappa/2$. (b) $P_{3H}$ versus $P_{\rm{in}}$. In these data, the pump frequency is tuned to maximize output power. The purple data correspond to $\Delta\nu \approx 0$, where $\Delta\nu$ is tuned to maximize output power. Blue and gold data correspond to $\Delta\nu=50\times\kappa/2$, where for blue data, $J$ is tuned to maximize output power, and $J=0$ for gold data. The y axes have units of photon number.}
    \label{fig:s1}
\end{figure*}
Here, we explain our methods to simulate THG, FWM-BS, DWE, and $\mu$OPO systems. For simplicity, we assume critically-coupled resonator modes, although the equations are easy to generalize. In the case of THG, we numerically integrate a set of coupled-mode equations (CMEs) that describe the evolution of intraresonator complex field variables $a$, $b_{\uparrow}$, and $b_{\downarrow}$, where $a$ denotes the pump field with angular frequency $\omega_{\rm{p}}$, and $b_{\uparrow (\downarrow)}$ denotes the generated third-harmonic field with angular frequency $3\omega_{\rm{p}}$ that co-propagates (counter-propagates) with the pump field. The CMEs are:
\begin{widetext}
\begin{equation}{\label{eq:CME}}
\begin{split}
    \frac{da}{dt} &= \sqrt{\frac{\kappa}{2\hbar \omega_{\rm{p}}}P_{\rm{in}}}-\left(\frac{\kappa}{2}-i2\pi\delta\right)a+ig_{\rm{0}}\left(|a|^2+2|b_{\uparrow}|^2+2|b_{\downarrow}|^2\right)a-3ig_{\rm{0}}a^{*2}b_{\uparrow}\\
    \frac{db_{\uparrow}}{dt} &=-\left(\frac{\kappa}{2}-i2\pi(\delta+\Delta\nu/2)\right)b_{\uparrow}+ig_{\rm{0}}\left(2|a|^2+|b_{\uparrow}|^2+2|b_{\downarrow}|^2\right)b_{\uparrow}-ig_{\rm{0}}a^3-iJb_{\downarrow}\\
    \frac{db_{\downarrow}}{dt} &=-\left(\frac{\kappa}{2}-i2\pi(\delta+\Delta\nu/2)\right)b_{\downarrow}+ig_{\rm{0}}\left(2|a|^2+2|b_{\uparrow}|^2+|b_{\downarrow}|^2\right)b_{\downarrow}-iJb_{\uparrow}
\end{split}
\end{equation}
\end{widetext}
where $\delta=(\omega_{\rm{0}}-\omega_{\rm{p}})/2\pi$ is the pump-resonator frequency detuning, $\Delta\nu=(\omega_{b}-3\omega_{\rm{p}})/2\pi$ is the frequency mismatch where $\omega_{\rm{b}}$ is the angular resonance frequency of the third-harmonic mode, and $g_{\rm{0}}$ is the single-photon nonlinear coupling (whose frequency dependence we neglect). Note that, in Eq. \ref{eq:CME}, $J$ has units rad/s, whereas it has units of Hz in the main text. In Fig. \ref{fig:s1}, we characterize our THG simulations. When $\Delta\nu=50\times\kappa/2$, the value of $J$ that maximizes the third harmonic power, $P_{3H}=|b_{\uparrow}|^2+|b_{\downarrow}|^2$, is not exactly $\Delta\nu$ due to self- and cross-phase modulation. Figure \ref{fig:s1}a shows $P_{\rm{3H}}$ versus $J$ for $P_{\rm{in}}=250$ $\mu$W. In Fig. \ref{fig:s1}b, we present simulated values of $P_{\rm{3H}}$ versus $P_{\rm{in}}$. For each data point, we tune $\omega_{\rm{p}}$ to maximize $P_{\rm{3H}}$. The data exhibit the expected cubic dependence of $P_{3H}$ on $P_{\rm{in}}$; when $P_{\rm{in}}$ becomes large, the conversion saturates. Intriguingly, we observe that, below saturation, $P_{\rm{3H}}$ is two times larger in non-photonic-crystal resonators where $\Delta\nu\approx 0$. However, in the saturation regime, the photonic crystal resonators generate the same $P_{\rm{3H}}$ values as the non-photonic-crystal resonators.

To analyze FWM-BS, DWE, and $\mu$OPO in photonic crystal resonators, we simulate two coupled LLE-type equations using the split-step Fourier method. The LLE is widely used to study microcombs because it encapsulates nonlinear interactions between many resonator modes using a single equation. Our coupled LLEs describe the evolution of CW and CCW intraresonator fields, denoted as $a_{\uparrow}$ and $a_{\downarrow}$, respectively. The equations are:  

\begin{equation}{\label{eq:LLE}}
\begin{split}
    \frac{da_{\uparrow, \downarrow}}{dt} &=\sqrt{\frac{\kappa}{2\hbar \omega_{\rm{p}}}P_{\rm{in}}}\left(1+\sum_{i}F_{i}e^{i(\Omega_{i}t-\mu_{i}\theta)}\right)\delta_{\uparrow}\\
    &-\left(\frac{\kappa}{2}+i\frac{\kappa}{2}\alpha\right)a_{\uparrow, \downarrow}+i\mathcal{D}(\mu)\tilde{a}_{\uparrow, \downarrow}-J(\mu)\tilde{a}_{\downarrow, \uparrow}\\
    &+ ig_{0}\left(|a_{\uparrow, \downarrow}|^2 + 2\int_{-\pi}^{\pi}\frac{|a_{\downarrow, \uparrow}|^2}{2\pi}d\theta\right)a_{\uparrow, \downarrow},
    \end{split}
\end{equation}

\noindent where $F_{i}$ is the amplitude, normalized to the primary pump laser amplitude, of the $i$th source (with frequency $\omega_{i}$) injected into resonator mode $\mu_{i}$ (relative to the pump mode). Hence, $\Omega_{i}=\mathcal{D}(\mu_{i})+\omega_{i}-\omega_{\rm{\mu}}+\frac{\kappa}{2}\alpha$, where $\alpha = \frac{2(\omega_{\rm{0}}-\omega_{\rm{p}})}{\kappa}$ is the normalized pump-resonator detuning, $\mathcal{D}(\mu)=\omega_{\rm{\mu}}-(\omega_{\rm{0}}+\mu D_{\rm{1}})$ is the integrated dispersion, where $D_{\rm{1}}=2\pi\times \textrm{FSR}$, $\tilde{a}_{\uparrow, \downarrow}$ indicates that operations are applied in the frequency domain, and $\theta$ is the azimuthal angle in a reference frame that moves at the group velocity (see Ref. \cite{taheri2017optical} for more details). The $\delta_{\uparrow}$ symbol indicates the driving terms are only applied to $a_{\uparrow}$. There are various approximations one can make to include cross-phase modulation (XPM) in Eq. \ref{eq:LLE}. Here, light travelling in each direction circulates the resonator many times in one simulation time step; therefore, we assess that XPM is suitably modeled using the averaged intraresonator intensities, $|a_{\uparrow, \downarrow}|^2$ (i.e., the final integral term in Eq. \ref{eq:LLE}).

Although the frequency mismatch, $\Delta \nu$, is not explicitly included in Eq. \ref{eq:LLE}, it is important to define this parameter in the case of FWM-BS, since its value dictates the required $J$. If the two pump modes have frequencies $\nu_{\rm{01}}$ and $\nu_{\rm{02}}$, the input mode has frequency $\nu_{\rm{in}}$, and the output mode has frequency $\nu_{\rm{out}}$, then $\Delta \nu=\nu_{\rm{01}}+\nu_{\rm{in}}-\nu_{\rm{02}}-\nu_{\rm{out}}$.

\section{Fabrication methods}

To create device layouts, we use the Nanolithography Toolbox, a free software package developed by NIST \cite{balram2016nanolithography}. We deposit stoichiometric SiN (Si$_3$N$_4$) by low-pressure chemical vapor deposition on top of a $3$ $\mu$m-thick layer of SiO$_{\rm{2}}$ on a $100$ mm diameter Si wafer. We fit ellipsometer measurements of the wavelength-dependent SiN refractive index and layer thicknesses to an extended Sellmeier model. The device pattern is created in positive-tone resist by electron-beam lithography and then transferred to SiN by reactive ion etching using a CF$_{\rm{4}}$/CHF$_{\rm{3}}$ chemistry. After cleaning the devices, we anneal them for four hours at $1100~^{\circ}$C  in N$_{\rm{2}}$. Next, we perform a liftoff of SiO$_2$ so that the resonator has an air top-cladding for dispersion purposes while the perimeter of the chip is SiO$_2$-clad for better coupling to lensed optical fibers. The facets of the chip are then polished for lensed-fiber coupling. After polishing, the chip is annealed again.

\bibliography{Bibliography}

\end{document}